# Dynamical investigation of macromolecular hybridization bioassays

R. Bittner, M. Wassermeier,

*Advalytix AG, Eugen Sänger Str. 53.0, D- 85649 Brunnthal, Germany*

*and*

A. Wixforth

*Experimentalphysik I, University of Augsburg, D- 86159 Augsburg, Germany*

Abstract

A novel sensoric technique for the dynamical in situ investigation of a hybridization bio assay is presented, which utilizes a metal bead labeling method. Therein, hybridization results in an increased metal coverage on parts of a substrate, where it takes place. Our sensing principle relies on the measurement of the radio frequency impedance of the hybridization spots. We propose several examples for sensor devices.

The hybridization of macromolecules like DNA or specific proteins for biochemical, medical or forensic purposes is usually observed applying fluorescence or radioactive markers. As the therefore required labeling processes are usually quite complicated, invasive, and also expensive, there is a growing need for detection of hybridization processes utilizing standardized, simple and cheap labeling. Moreover, standard hybridization assays usually only generate so called `end point´ results, i.e. the kinetics of the hybridization process cannot be studied.

Apart from using fluorescent dyes or radioactive markers, the application of nano-scale gold-beads for labeling purposes has also been reported upon [1]. On the one hand, the gold-beads can be used for labeling the sample molecule, which binds specifically to its counterpart on a micro-array. On the other hand, a two part sandwich assay can be realized, which consists of a defined binding site for the sample molecule, a labeled head group in solution, and the sample molecule acting as a specific linker between the both [2,3].

Fig. 1 depicts the idea behind these metal labeling techniques. In the simplest arrangement e.g. single stranded and gold-labeled oligonucleotides bind directly and specifically to the probe molecules having the complementary sequence (Fig. 1a). Fig. 1b illustrates the aforementioned two step sandwich technique. The probe molecules on the chip surface provide the complementary sequence for a moiety of the sample molecule, another moiety of which specifically binds to a third molecule being gold-labeled.

(a) (b)

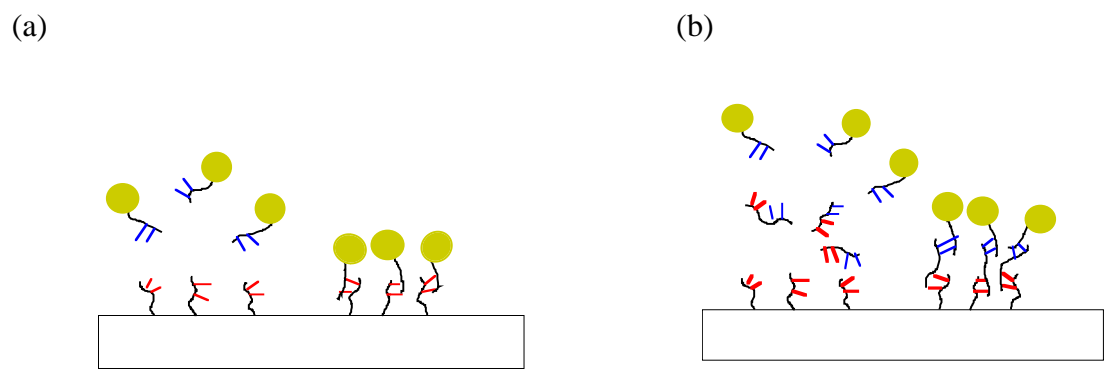

**Fig. 1.:** Schematic representation of a gold bead labeling assay for hybridization experiments. (a): two species assay, with the probe molecules attached to an appropriate substrate and the sample molecules in solution labeled with gold beads. Hybridization yields a specific binding of the labeled sample molecules to the probe sites. (b): two step sandwich technique, where the sample molecules themselves are not labeled. Instead, they partially link to probe sites attached to the substrate as well as a third species being gold labeled. Thus, also in this case hybridization yields a gold bead specifically attached to the substrate surface.

A positive hybridization process can than be probed by several experimental techniques, among of which are micro-gravimetry [4,5,6], cyclo-voltammetry [7,8], and optical methods like ellipsometry [9,10] or surface plasmon resonances [11,12].

Recently, a purely electrical method for the detection of a gold labelled hybridization assay has been reported [13,14]. The two step sandwich technique was applied to mark a positive hybridization process on a defined spot of a micro array. After hybridization and washing, the gold-beads attached to the substrate surface were additionally metalized electrochemically (in this case using Ag) in order to increase their size until a closed current path was formed between two micro electrodes. Positive hybridization was thus indicated by metallic conductivity between the electrodes. However, also this experiment yields only an end point result and does not allow to observe the hybridization kinetics.

Here, we propose a novel experimental technique, which allows for the direct observation of the hybridization dynamics in DNA or protein bioassays. Basically, we also rely on the aforementioned gold bead marker techniques (cf. Fig. 1). However, in contrast to the experiments outlined before, we determine the real gold coverage on specific binding sites directly by means of high-frequency impedance determination of the reaction spot on a micro array. This way, not only end point results are obtained but rather the kinetics of the hybridization process can be monitored as well with high accuracy. An integrated micro fluidic system 'on the chip' enables us to create a sophisticated sensor element for many different bioassay applications.

The principle of our experiment is sketched in Fig. 2. The increasing gold surface coverage on a positive spot is detected by measuring the high frequency (RF) impedance of the sensing part of the chip. In the simplest case, the active sensor element is an impedance matched pair or array of high frequency electrodes close to the chip surface. This electrode pair or array might be part of a RF transmission line. In this case, the transmission and reflection behavior of the transmission line will strongly depend on the degree of the gold coverage close to the surface of the electrode array. In detail, capacitive coupling between the metal particles provides the dominant contribution to the electrical modification of the RF impedance.

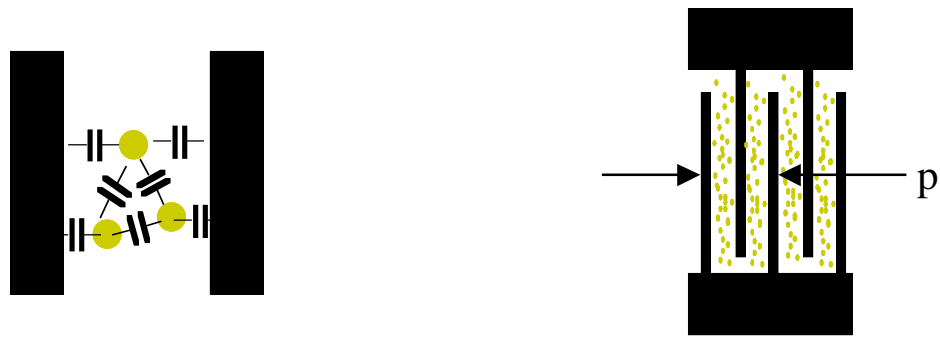

**Fig. 2:** Schematic sketch of an electrode structure for determination of the RF impedance within an hybridization spot. The surface bound gold beads couple primarily capacitively to the electrodes. Hence, the RF impedance of the latter depends on the degree of the gold coating upon the active area. p depicts the finger periodicity of the electrode structure.

The gold layer indicating a positive hybridization is at maximum a monolayer of gold particles very close to the sensor surface. Hence, either a special field distribution between the electrodes or a dominantly surface sensitive measuring scheme is required. A well established surface sensitive method for such a purpose is the application of surface acoustic waves (SAW's). SAW's are modes of elastic energy propagating along the surface of a solid substrate. If the substrate is piezoelectric, both the excitation as well as the detection of SAW is particularly simple [15]. Transducers consisting of two comb-like interdigitated metal structures (Fig. 2, right) can be used, which convert an appropriate RF voltage signal into a surface acoustic wave. The frequency $f$ of the applied RF signal must meet the zero order resonance condition

$$f = v_{SAW} \cdot p \tag{1}$$

in order to operate such a so called interdigital transducer (IDT). $v_{SAW}$ denotes the SAW velocity and $p$ is the finger periodicity of the IDT (see Fig. 2). Please note, that there are higher order resonance conditions, which are not considered here. SAW's on piezoelectric substrates are most favorable for our purpose, i.e. detection of gold labeled macromolecular hybridization processes. We point out, that SAW's on piezoelectric substrates can be used to probe RF impedance changes of the sensor electrodes as depicted in Fig. 2, as well as to probe the RF impedance of the gold covered surface itself .

We propose a practical example of a very sensitive detector for thin gold layers based on the well known feature of SAW's to be reflected by periodic structures inside the propagation path of the SAW (hereafter simply referred to as "sound path"). In general, a SAW is reflected (and/or scattered) by any surface inhomogenity, which either changes the mechanical or the electrical boundary conditions for the SAW. In practical application for reflecting structures, the mechanical boundary conditions are typically changed by etching appropriately oriented grooves in to the substrate's surface. In order to change the electrical boundary conditions periodic arrays of electrically well conducting metal stripes have been widely applied. In particular the latter case is relevant for sensoric applications, as the reflectivity of a metal stripe array strongly depends on its electrical impedance. Fig. 3 depicts the simplest arrangement for a SAW based hybridization sensor utilizing a metallic reflector array. A SAW is generated by the IDT structure, which is driven by a RF pulse, and propagates to the right until it enters the reflector element REFL. The hybridization spot is localized atop of the reflector element as indicated by the gray area. We point out, that the droplet containing probe and marker molecules must to be outside the sound path if Rayleigh waves are used, since the latter are heavily attenuated by viscous material at the surface due to mechanical damping of the displacement component in the sagittal plane.

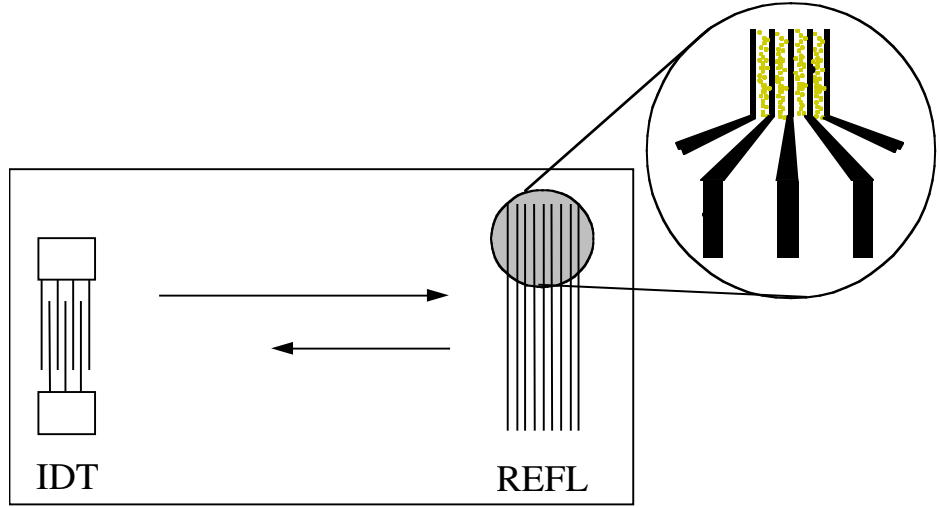

**Fig. 3:** Simple representation of a SAW based hybridization sensor. A SAW is generated by the interdigital transducer (IDT). After reflection at the reflector array (REFL), the SAW can be analyzed in terms of amplitude and phase at the same transducer. The reflection coefficient of the reflector array strongly depends on its load impedance, which, in turn, depends on its gold-bead coverage. The ‚biological' sampling area (spot) is located outside the sound path, which enables the application of Rayleigh-type SAW's.

As described above, the gold beads close to the substrate surface capacitively couple adjacent fingers of the reflector array (see Fig. 2) and, therewith, change its complex termination impedance $Z_{load}$. According to the well known P-matrix formalism , the acoustic reflectivity of a reflector array carrying a complex termination impedance is given by:

$$P_{11}(Z_{load}) = P_{11}^{sc} + \frac{2 \cdot P_{13}^2}{P_{33} + \frac{1}{Z_{load}}} \qquad (2)$$

where $P_{11}^{sc}$ is the short circuit acoustic reflectivity. For a split-finger reflector array it is approximately zero. $P_{13}$ and $P_{33}$ are the reflector's P-matrix elements, which depend on its geometry and the material of the reflector fingers [16,17].

We point out, that also a series of reflector elements as described above may be used in order to investigate the hybridization processes in a manifold of different spots simultaneously. Each reflector array then generates its own ‚echo', which can be analyzed selectively. However, performing the experiment only time resolved as outlined before bears the problem of multiple echoes from the manifold of reflector elements, the unambiguous separation of which is difficult. This problem can be overcome, if the experiment is carried out additionally frequency resolved using some kind of broadband IDT's (e.g. tapered structures [18]) and reflector elements with different periodicity. A principal scheme of this layout is depicted in Fig. 4. Thus, the sensor principle we propose is supposed to be applicable for high-throughput screening purposes as well.

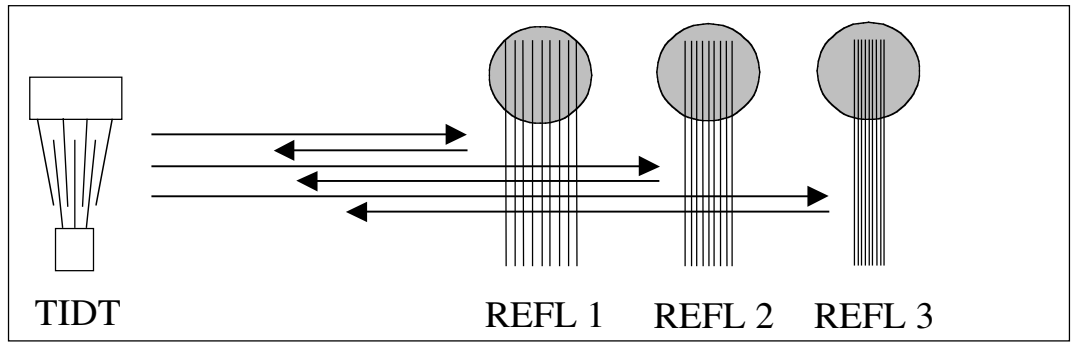

**Fig. 4:** Illustration of a SAW based hybridization sensor for high-throughput screening. SAW's of different wavelength are generated by means of a broadband interdigital transducer (here a tapered structure "TIDT" is shown exemplarily). After reflection at the different reflector arrays (REFL 1 to 3), the SAW's of different frequency can each be analyzed independently in terms of amplitude and phase at the same transducer.

The simple layouts as depicted in Figs. 3 and 4, however, inherently suffer from the basic need for time resolution in order to separate the reflected SAW-signals from the RF pulse used to initially generate the SAW, an experimental technique which may become quite cumbersome. Therefore, we designed an autocorrelation scheme (Fig. 5), which makes time resolution redundant:

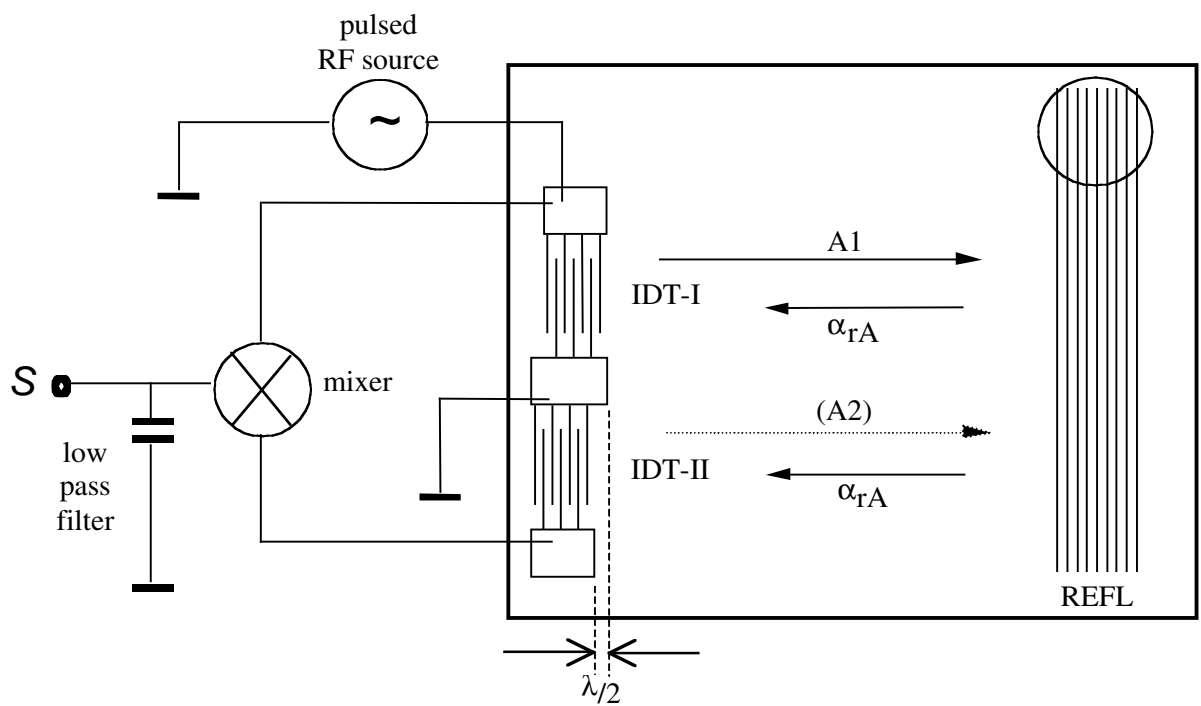

**Fig. 5:** Illustration of the autocorrelation sensor as described in the text. A1 and A2 are the primary and the secondary sound path, respectively. The initial SAW propagates only along A1, however, the reflected SAW propagates along the whole aperture of the reflector elements, i.e. along A1 as well as along A2. Note, that IDT-I and IDT-II are laterally displaced by $\lambda/2$ in order to maximize the autocorrelation signal as discussed in the text.

This layout uses two independent interdigital transducers, IDT-I and IDT-II, and a reflector element. The hybridization sample spot is located outside the sound path as discussed before. A RF signal is fed into IDT-I generating a surface acoustic wave of amplitude A, which propagates along the sound path A1 towards the reflector array. At the reflector element with reflectivity *R*, a part of the incident SAW intensity, $\alpha RA$, is reflected back towards IDT-I. However, the reflector array is now larger than the aperture of IDT-I covering not only the primary sound path A1 but also the secondary sound path A2 and emits over its total aperture. Accordingly, also IDT-II is penetrated by the reflected SAW. Therefore, after the time of flight given by twice of the single length of the sound paths as well as the sound velocity of the substrate, both transducers, IDT-I and IDT-II, receive a signal, which is proportional to the reflectivity of the reflector. As discussed before, the latter depends on the load impedance of the whole reflector array.

The signals received by both IDT's can now be mixed homodynely (i.e. be multiplied) by means of, for instance, a double balanced mixer. This way, an output signal is obtained consisting of two components, the one having the sum frequency ($\omega_1 + \omega_2 = 2\omega$) and at the other having the difference frequency ($\omega_1 - \omega_2 = 0$). The first can be suppressed by a low pass filter and eventually solely the DC component *S* of the product of the particular signals is obtained as actual measurement signal.

Please note, that *S* depends on the phase difference $\Delta\phi$ between the particular input signals according to:

$$S = (\alpha \cdot r \cdot A)^2 \cos(\Delta\phi). \tag{3}$$

Hence, *S* can be maximized by displacing IDT-I and IDT-II by, e.g., $\lambda/2$ laterally with respect to each other as indicated in Fig. 5.

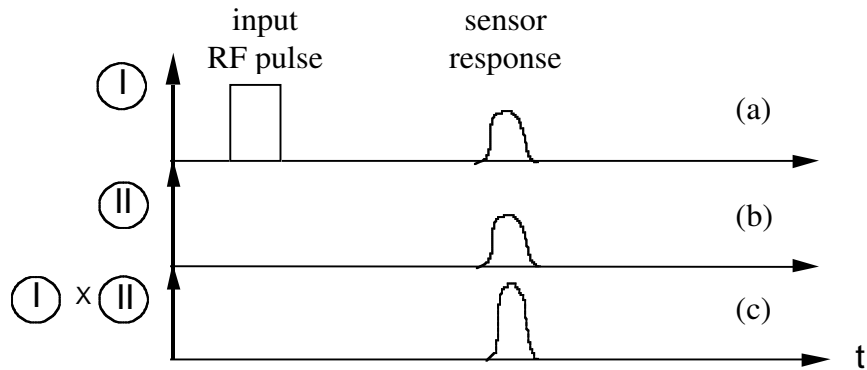

**Fig. 6:** General response scheme in the time domain as will be obtained from the autocorrelation sensor depicted in Fig. 5. (a) shows the signals occurring at IDT-I. (b) shows the signals occurring at IDT-II. (c) shows the autocorrelation signal.

Fig. 6 depicts the general response scheme, which will be obtained from an autocorrelation experiment as outlined before. The sensor response signals obtained from the particular transducers IDT-I and IDT-II are indicated together with the result of their analog multiplication followed by a low pass filter. The mixer output signal *S* is only non-zero for the reflected SAW pulse, as indicated in Fig. 6(c).

The sensor system we propose in this paper is extremely fast as compared to the typical kinetics of the hybridization process, since the delay-times between the RF input pulse and the reflected SAW pulse of our SAW sensors are at the order of microseconds. Thus, our sensor systems allows for the examination of the kinetics of hybridization process with high time resolution, which is vitally important for many bio assays. Moreover, in combination with a SAW based micro fluidic platform [19], very complex and sophisticated bio assays can be realized on a single chip.

A manifold of different schemes for SAW based hybridization sensors may be taken into consideration: For instance, the reflector array of our sensor might serve as a probe head for a separate, external sensor element, which then may directly be inserted into a micro titer well. Furthermore, the fundamental concept of our sensor is not restricted to SAW reflectors. Alternatively, resonator structures may be used, the frequency response and the bandwidth of which are similarly affected by a capacitive load between the resonator elements. Another way to realize SAW based sensors utilizing the concepts proposed in this paper arises from the application of multi-strip couplers, the coupling efficiency of which also depends on the impedance of the coupler element. These variations of the presented sensor concept are presently under evaluation in our group.

A different approach to SAW based hybridization sensors arises from the fact that IDTs for SAW applications can be tailored in terms of their frequency response. In zero order approximation, the design of an interdigital transducer (i.e. the finger period, the metallization ratio, the numbers of the fingers, and the overlap between adjacent fingers etc.) is coorelated with the frequency response of the IDT by a simple Fourier transformation [20]. Hence, capacitive coupling of the IDT fingers via gold beads attached to the substrate's surface will transform the information of the status of the hybridization assay from the time domain into the frequency domain. This approach, too, is presently investigated in our group and details will be reported elsewhere.

We point out that our sensors may even be operated wireless, since typical operating frequencies for SAW devices are at the order of 100 MHz up to 1 GHz. For wireless operation, an input signal having the operating frequency is sent to the SAW device by means of a transmitter. Subsequently, after the delay time of the SAW device has passed, the device will respond a signal containing the desired information, which then can be analyzed in a receiving unit.

Finally note, that the physically most simple realization of our sensor concept is represented by a direct determination of the complex impedance of an IDT structure on an arbitrary insulating surface. However, although being physically simple, this direct approach would require much more sophisticated measurement techniques, which renders this way uneconomical.

In conclusion, we propose a novel sensor concept for the in situ observation of hybridization bio assays. It is based on a metal bead marker method yielding capacitive coupling of different electrodes on an appropriate surface. Applying this principle, we have proposed various surface-sensitive sensor systems based on acoustic surface wave devices. By means of an autocorrelation setup, which can be realized ‚on chip', we were able to design a simple experiment, which allows the in-situ examination of the kinetics of the hybridization process.

We gratefully acknowledge very useful discussions with C. Gauer, J. Scriba, M. Sickert, and A. Rathgeber of Advalytix, and financial support of the same company.